\documentclass[a4paper,12pt]{article}
\usepackage{jheppub}
\usepackage{epsfig}
\usepackage[T1]{fontenc}
\usepackage{amsmath}
\usepackage{amssymb}
\usepackage{graphicx}
\usepackage{amsfonts}
\usepackage{hyperref}

\title{A Model of Radiative Neutrino Mass: \\with or without Dark Matter}
\author[a,b]{Amine Ahriche}\author[c]{Kristian~L. McDonald}\author[d]{and Salah Nasri}
\affiliation[a]{Department of Physics, University of Jijel, PB 98
Ouled Aissa, DZ-18000 Jijel, Algeria} \affiliation[b]{The Abdus
Salam International Centre for Theoretical Physics, Strada Costiera
11, I-34014, Trieste, Italy} \affiliation[c]{ARC Centre of Excellence for Particle Physics at the Terascale,\\
School of Physics, The University of Sydney, NSW 2006, Australia}
\affiliation[d]{Department of Physics, UAE University, P.O. Box
17551, Al-Ain, United Arab Emirates} \emailAdd{aahriche@ictp.it,
klmcd@physics.usyd.edu.au, snasri@uaeu.ae.ac}

\abstract{We present a three-loop model of neutrino mass whose
most-general Lagrangian possesses a softly-broken accidental $Z_2$
symmetry. In the limit that a single parameter vanishes,
$\lambda\rightarrow0$, the $Z_2$ symmetry becomes exact and the
model contains a stable dark-matter candidate. However, even for
finite $\lambda\ll1$, long-lived dark matter is possible, giving a
unified solution to the neutrino mass and dark matter problems that
does not invoke a new symmetry. Taken purely as a neutrino mass
model, the new physics can be at the TeV scale. When dark matter is
incorporated, however, only a singlet scalar can remain this light,
though the dark matter can be tested in direct-detection
experiments.} 

\begin{document}

\maketitle

\section{Introduction\label{sec:introduction}}

The observation of neutrino oscillations in solar, atmospheric and
reactor experiments confirms that neutrinos are massive and that the
Standard Model (SM) is incomplete. Another strong motivation for
beyond-SM physics comes from astrophysical observations, which
motivate a new gravitating particle species referred to as dark
matter (DM). It is sensible to ask if these two problems could have
a common solution. Models with radiative neutrino
mass~\cite{Zee:1980ai} offer a promising direction for a unified
solution to these problems (for a discussion of radiative models see
e.g.~\cite{Angel:2012ug}). If the coupling to DM is related to the
source of lepton number symmetry breaking, DM can propagate inside
the loop diagram that generates neutrino mass, killing the
proverbial two birds with one stone.

An early proposal along these lines was put forward by Krauss, Nasri
and Trodden (KNT)~\cite{Krauss:2002px} (for analysis see
Refs.~\cite{Baltz:2002we,Cheung:2004xm,Ahriche:2013zwa,Ahriche:2014cda}).
In this paper we investigate a three-loop model of neutrino mass
that is related to the KNT model. The model differs in its field
content but employs a three-loop diagram with the same topology. The
use of distinct beyond-SM multiplets produces some key differences.
Recall that the KNT model utilizes a discrete ($Z_2$) symmetry,
which serves two purposes: It precludes tree-level neutrino mass,
which would otherwise dominate the loop mass, and it gives a stable
particle that is taken as the DM. Consequently, the KNT model gives
a unified solution to the neutrino mass and DM problems.

Different from the KNT model, the present model does not require a new
symmetry to preclude tree-level neutrino mass, despite sharing the same
loop-topology. It is therefore a viable model of radiative neutrino mass
independent of any DM considerations. Interestingly, the most-general
Lagrangian for the model possesses a softly-broken accidental $Z_2$
symmetry. In the limit where a single parameter vanishes, $%
\lambda\rightarrow0$, this symmetry becomes exact and the model
contains a stable DM candidate. As a result, the DM width goes like
$\Gamma_{DM}\propto \lambda^2$ for nonzero $\lambda$, and one can
always make this sufficiently small to obtain long-lived DM, or
simply take $\lambda\rightarrow0$ for absolutely stable DM. Thus, DM
is possible with or without the $Z_2$ symmetry. This gives a unified
solution to the DM and neutrino mass problems that does not require
a new symmetry. Importantly, the limit $\lambda\rightarrow0$ does
not affect the predictions for neutrino mass. The $Z_2$ symmetry is
essentially the same one found in the KNT model (and the related
triplet model~\cite{Ahriche:2014cda}), though in those cases the
most-general Lagrangian contains multiple symmetry breaking terms,
including ones that give tree-level neutrino mass.

We shall see that the phenomenology of the model depends on the region of
parameter space considered. Taken purely as a model of neutrino mass, the
new physics can be at the TeV scale and may be probed in collider
experiments. When DM is incorporated one requires $M_{DM}\sim10$~TeV,
putting some of the new multiplets beyond the reach of colliders. None the
less, a singly-charged scalar that appears in the model can remain at the
TeV scale, with or without the inclusion of DM. Even when DM is included,
prospects for testing the model in direct-detection experiments are good.

We note that one-loop models of neutrino mass that admit DM candidates but
do not require a symmetry to exclude tree-level masses exist \cite%
{Law:2013saa}, with one model further studied in Ref.~\cite{Brdar:2013iea}.
Other works studying connections between neutrino mass and DM include Refs.~%
\cite%
{Ma:2006km,Kajiyama:2013zla,Kanemura:2011mw,Ho:2013hia,MarchRussell:2009aq}.
For a review see \cite{BMV}.

The layout of this paper is as follows. In Section~\ref{sec:model5} we
describe the basic details of the model. Neutrino masses are calculated in
Section~\ref{sec:nuetrino_mass5} and important flavor-changing constraints
are discussed in Section~\ref{sec:constraints5}. We consider DM in Section~%
\ref{sec:dark_matter5}, discussing the issue of longevity and the relic
abundance. Our main numerical results and discussion are given in Section~%
\ref{sec:results} and we comment on collider phenomenology in Section~\ref%
{sec:collider}. We briefly describe interesting generalizations of our model
in Section~\ref{sec:generalize_KNT}, and conclude in Section~\ref{sec:conc5}.

\section{The Model\label{sec:model5}}

We extend the SM to include a charged scalar singlet, $S^{+}\sim (1,1,2)$, a
complex scalar quintuplet, $\phi \sim (1,5,2)$, and a real fermion
quintuplet, $\mathcal{F}\sim (1,5,0)$. We write the exotics in
symmetric-matrix form as $\phi _{abcd}$ and $\mathcal{F}_{abcd}$, where
\begin{eqnarray}
&&\ \phi _{1111}=\phi ^{+++},\ \phi _{1112}=\frac{\phi ^{++}}{\sqrt{4}},\
\phi _{1122}=\frac{\phi ^{+}}{\sqrt{6}},\ \phi _{1222}=\frac{\phi ^{0}}{%
\sqrt{4}},\ \phi _{2222}=\phi ^{-}, \\
&&\ \mathcal{F}_{1111}=\mathcal{F}_{L}^{++},\ \mathcal{F}_{1112}=\frac{%
\mathcal{F}_{L}^{+}}{\sqrt{4}},\ \mathcal{F}_{1122}=\frac{\mathcal{F}_{L}^{0}%
}{\sqrt{6}},\ \mathcal{F}_{1222}=\frac{(\mathcal{F}_{R}^{+})^{c}}{\sqrt{4}}%
,\ \mathcal{F}_{2222}=(\mathcal{F}_{R}^{++})^{c}.  \notag
\end{eqnarray}%
Note that $\phi ^{+}$ and $\phi ^{-}$ are distinct fields and, in
particular, $\phi ^{-}\neq (\phi ^{+})^{\ast }$. The Lagrangian for the
model contains the terms
\begin{equation}
\mathcal{L}\supset \mathcal{L}_{\text{{\tiny SM}}}+\{f_{\alpha \beta }\,%
\overline{L_{\alpha }^{c}}\,L_{\beta }\,S^{+}+g_{i\alpha }\,\overline{%
\mathcal{F}_{i}}\,\phi \,e_{\alpha R}+\mathrm{H.c}\}\;-\;\frac{1}{2}\,%
\overline{\mathcal{F}_{i}^{c}}\,\mathcal{M}_{ij}\,\mathcal{F}%
_{j}\;-\;V(H,S,\phi ).
\end{equation}%
We label lepton flavors by lower-case Greek letters, $\alpha ,\,\beta \in
\{e,\,\mu ,\,\tau \}$, while exotic fermion generations are labeled by $i$.
The superscript \textquotedblleft $c$" denotes charge conjugation. The
Lagrangian shows that the multiplets $\phi $ and $\mathcal{F}$ are
sequestered from the SM neutrinos. None the less, they play a key role in
enabling neutrino mass, as we shall shortly see.

The explicit expansion of the fermion mass term gives
\begin{eqnarray}
&&-\frac{1}{2}\,(\overline{\mathcal{F}_{i}^{c}})_{abcd}\,\mathcal{M}_{ij}\,(%
\mathcal{F}_{j})_{efgh}\,\epsilon ^{ae}\,\epsilon ^{bf}\,\epsilon
^{cg}\,\epsilon ^{dh}+\mathrm{H.c.}  \notag \\
&=&-\overline{\mathcal{F}_{iR}^{++}}\,\mathcal{M}_{ij}\,\mathcal{F}%
_{jL}^{++}+\overline{\mathcal{F}_{iR}^{+}}\,\mathcal{M}_{ij}\,\mathcal{F}%
_{jL}^{+}-\frac{1}{2}\overline{(\mathcal{F}_{iL}^{0})^{c}}\,\mathcal{M}%
_{ij}\,\mathcal{F}_{jL}^{0}+\mathrm{H.c.}  \notag \\
&=&-\overline{\mathcal{F}_{i}^{++}}\,\mathcal{M}_{ij}\,\mathcal{F}_{j}^{++}-%
\overline{\mathcal{F}_{i}^{+}}\,\mathcal{M}_{ij}\,\mathcal{F}_{j}^{+}-\frac{1%
}{2}\overline{\mathcal{F}_{i}^{0}}\,\mathcal{M}_{ij}\,\mathcal{F}_{j}^{0},
\end{eqnarray}
where, in the last line, we define:
\begin{equation}
\mathcal{F}^{++}=\mathcal{F}_{L}^{++}+\mathcal{F}_{R}^{++}\,,\quad \mathcal{F%
}^{+}=\mathcal{F}_{L}^{+}-\mathcal{F}_{R}^{+}\,,\quad \mathcal{F}^{0}=%
\mathcal{F}_{L}^{0}+(\mathcal{F}_{L}^{0})^{c}.
\end{equation}
Here $\mathcal{F}^{0}$ is clearly a Majorana fermion, while the other four
components of $\mathcal{F}$ partner-up to give two charged (Dirac) fermions.
Without loss of generality we work in a basis with $\mathcal{M}_{ij}=\mathrm{%
diag}(M_{1},\,M_{2},\,M_{3})$, where the masses are ordered as $%
M_{1}<M_{2}<M_{3}$. In what follows, we will use $M_{F}\equiv M_{1}$ for the
DM mass.

In terms of these fields the Yukawa couplings involving the new fermions are
written as
\begin{eqnarray}
&&g_{i\alpha }\,(\overline{\mathcal{F}_{i}})^{abcd}\,\phi _{abcd}\,e_{\alpha
R}\ =\ g_{i\alpha }\,\left\{ \phi ^{+++}\overline{\mathcal{F}_{i}^{++}}%
\,P_{R}\,e_{\alpha }+\phi ^{++}\overline{\mathcal{F}_{i}^{+}}%
\,P_{R}\,e_{\alpha }\right.  \notag \\
&&\,\left. +\phi ^{+}\overline{\mathcal{F}_{i}^{0}}\,P_{R}\,e_{\alpha }-\phi
^{0}\,\overline{(e_{\alpha })^{c}}\,P_{R}\mathcal{F}_{i}^{+}\,+\phi ^{-}%
\overline{(e_{\alpha })^{c}}\,P_{R}\,\mathcal{F}_{i}^{++}\right\} ,
\end{eqnarray}%
where $P_{R}$ is a standard projection operator. The extra minus sign is due
to the definition of $\mathcal{F}^{+}$.

We consider the parameter space where the SM Higgs breaks the electroweak
symmetry via the nonzero vacuum value $\langle H\rangle \neq 0$, while $%
\langle \phi \rangle =0$, so the SM tree-level value of the $\rho $%
-parameter is not modified. Before turning to neutrino mass we would like to
discuss a few features of the model. To this end, let us briefly consider
the theory in the absence of the singlet $S$. In this case the scalar
potential is
\begin{equation}
V(H,\,\phi )=V(H)+V(\phi )+V_{m}(H,\,\phi ),
\end{equation}%
where the mixing potential is
\begin{equation}
V_{m}(H,\,\phi )=\lambda _{H\phi 1}(\phi ^{\ast })^{abcd}\phi
_{abcd}(H^{\ast })^{e}H_{e}+\lambda _{H\phi 2}(\phi ^{\ast })^{abcd}\phi
_{ebcd}(H^{\ast })^{e}H_{a}.
\end{equation}%
This potential $V(H,\,\phi )$ possesses an accidental $U(1)$ symmetry, $\phi
\rightarrow e^{i\theta }\phi $. However, the coupling to $\mathcal{F}$
breaks this symmetry to a discrete subgroup, due to the Majorana mass.
Therefore in the absence of $S$, the theory has an accidental $Z_{2}$
symmetry:
\begin{equation}
\{\phi ,\,\mathcal{F}\}\rightarrow \{-\phi ,\,-\mathcal{F}\}.
\end{equation}%
Adding $S$ to the theory, the full potential can be written as
\begin{equation}
V(H,\,S,\,\phi )=V(H,\,\phi )+V(S)+V_{m}(S,\,\phi
)+V_{m}(H,\,S)+V_{m}(H,\,S,\,\phi ),
\end{equation}%
where the first four terms in this potential all preserve the discrete
symmetry, and the last mixing-potential is given by
\begin{equation}
V_{m}(H,\,S,\,\phi )=\frac{\lambda _{S}}{4}(S^{-})^{2}\phi _{abcd}\phi
_{efgh}\epsilon ^{ae}\epsilon ^{bf}\epsilon ^{cg}\epsilon ^{dh}+\lambda
S^{-}(\phi ^{\ast })^{abcd}\phi _{abef}\phi _{cdjl}\epsilon ^{ej}\epsilon
^{fl}+\mathrm{H.c.}
\end{equation}%
The first term in this potential also preserves the $Z_{2}$ symmetry,
leaving the second term as the sole source of $Z_{2}$ symmetry-breaking in
the full theory. Thus, in the limit $\lambda \rightarrow 0$ the theory
possesses the $Z_{2}$ symmetry $\{\phi ,\,\mathcal{F}\}\rightarrow \{-\phi
,\,-\mathcal{F}\}$, making $\lambda \ll 1$ technically natural. This
symmetry is analogous to that invoked in both the KNT model~\cite%
{Krauss:2002px} and the three-loop model with triplets~\cite{Ahriche:2014cda}%
. In the limit that $\lambda \rightarrow 0$ a stable particle emerges, which
we return to in Section~\ref{sec:dark_matter5}.

If the $Z_2$ symmetry were exact, it would prevent mixing between $\mathcal{F%
}$ and the SM leptons. Consequently any such mixing must be generated
radiatively and must involve the coupling $\lambda$. This mixing is of a
sufficiently high order as to be negligible, though to be certain one can
always choose $\lambda$ sufficiently small to make the mixing negligible. We
can therefore ignore any mixing between $\mathcal{F}$ and the SM.

At tree-level the components of $\mathcal{F}$ are mass-degenerate, while the
components of $\phi$ experience a mild splitting due to the $%
\lambda_{H\phi2} $-term in $V_m(H,\,\phi)$. For $M_\phi\gtrsim\mathcal{O}(%
\mathrm{TeV})$ this mass-splitting is not significant and is negligible for $%
\lambda_{H\phi2}\lesssim0.1$. Thus, to good approximation the components of $%
\mathcal{F}$ and $\phi$ are degenerate at tree-level, with masses $M_%
\mathcal{F}$ and $M_\phi$, respectively. Radiative corrections lift these
mass degeneracies. For example, loops involving SM gauge bosons induce
splittings of $M_{\mathcal{F}^{++}}-M_{\mathcal{F}^+}\simeq 490$~MeV, and $%
M_{\mathcal{F}^+}-M_{\mathcal{F}^0}\simeq163$~MeV, among the components of $%
\mathcal{F}$, leaving $\mathcal{F}^0$ as the lightest state once
loop-corrections are incorporated~\cite{Cirelli:2005uq,Kumericki:2012bh}.
Similar splittings are induced for the components of $\phi$~\cite%
{Cirelli:2005uq}. For most purposes these small splittings can be neglected.

We note that the fermions $\mathcal{F}\sim(1,5,0)$ employed in this model
were studied in a number of other contexts. They allow a generalization of
the Type-III seesaw mechanism~\cite{Foot:1988aq} that achieves neutrino mass
via a low-energy effective operator of mass-dimension $d=9$~\cite%
{Kumericki:2012bh,Liao:2010cc,McDonald:2013kca}. Similarly they permit a
generalized inverse seesaw mechanism~\cite{Law:2013gma}. The neutral
component of the fermion is also the favored "Minimal DM" candidate~\cite%
{Cirelli:2005uq}. For related phenomenological studies see Refs.~\cite%
{Chen:2013xpa,Ding:2014nga}.

\section{Three-Loop Radiative Neutrino Masses\label{sec:nuetrino_mass5}}

\begin{figure}[t]
\begin{center}
\includegraphics[width = 0.5\textwidth]{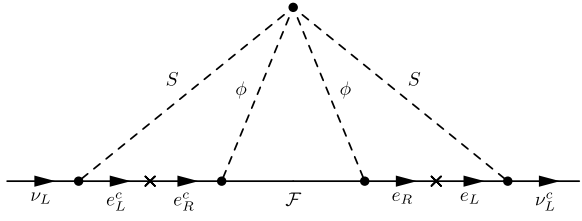}
\end{center}
\caption{Three-loop diagram for radiative neutrino mass, where $S$ and $%
\phi$ are new scalars and $\mathcal{F}$ is an exotic fermion.}
\label{fig:3loop_nuDM5}
\end{figure}

The Yukawa Lagrangian is not sufficient to break lepton number symmetry.
However, as just mentioned, the scalar potential contains the terms
\begin{eqnarray}
V(H,S,\phi ) &\supset &\frac{\lambda _{S}}{4}(S^{-})^{2}\phi _{abcd}\phi
_{efgh}\epsilon ^{ae}\epsilon ^{bf}\epsilon ^{cg}\epsilon ^{dh}+\frac{%
\lambda _{S}^{\ast }}{4}(S^{+})^{2}(\phi ^{\ast })^{abcd}(\phi ^{\ast
})^{efgh}\epsilon _{ae}\epsilon _{bf}\epsilon _{cg}\epsilon _{dh}  \notag \\
&=&\frac{\lambda _{S}}{2}(S^{-})^{2}\{\phi ^{+++}\phi ^{-}-\phi ^{++}\phi
^{0}+\frac{1}{2}\phi ^{+}\phi ^{+}\}+\mathrm{H.c.}
\end{eqnarray}%
When combined with the Yukawa couplings, these ensure that lepton number
symmetry is explicitly broken in the model. Consequently, Majorana neutrino
masses are generated radiatively, appearing at the three-loop level as shown
in Figure~\ref{fig:3loop_nuDM5}. The are actually five distinct diagrams,
corresponding to the sets $\{\phi ^{+},\mathcal{F}^{0},(\phi ^{+})^{\ast }\}$%
, $\{\phi ^{++},(\mathcal{F}^{+})^{c},\phi ^{0}\}$, $\{\phi ^{0},\mathcal{F}%
^{+},\phi ^{--}\}$, $\{\phi ^{+++},(\mathcal{F}^{+})^{c},(\phi ^{-})^{\ast
}\}$ and $\{\phi ^{-},\mathcal{F}^{++},\phi ^{---}\}$ propagating in the
inner loop in Figure \ref{fig:3loop_nuDM5}.

In the limit where the mass-splitting among components of $\phi $ and $%
\mathcal{F}$ are neglected, the loop-diagrams gives
\begin{equation}
(\mathcal{M}_{\nu })_{\alpha \beta }=\frac{5\lambda _{S}}{(4\pi ^{2})^{3}}%
\frac{m_{\gamma }m_{\delta }}{M_{\phi }}\,f_{\alpha \gamma }\,f_{\beta
\delta }\,g_{\gamma i}^{\ast }\,g_{\delta i}^{\ast }\times F\left( \frac{%
M_{i}^{2}}{M_{\phi }^{2}},\frac{M_{S}^{2}}{M_{\phi }^{2}}\right) .
\end{equation}%
Here the function $F$ encodes the loop integrals (see Appendix~\ref%
{app:loop_integral}) and has the same form as given in Ref.~\cite%
{Ahriche:2013zwa}. $M_{S}$ is the charged-singlet mass and $M_{\phi}$ is the
mass of the degenerate members of $\phi $.

The entries in the neutrino mass matrix $(\mathcal{M}_{\nu})_{\alpha \beta }
$ may be related to the mass eigenvalues and the elements of the
Pontecorvo-Maki-Nakawaga-Sakata (PMNS) mixing matrix~\cite{Pontecorvo:1967fh}%
:
\begin{equation}
(\mathcal{M}_{\nu })_{\alpha \beta }=[U_{\nu }\cdot \mathrm{diag}%
(m_{1},\,m_{2},\,m_{3})\cdot U_{\nu }^{\dag }]_{\alpha \beta }.
\end{equation}%
We parameterize the PMNS matrix as
\begin{equation}
U_{\nu }=\left(
\begin{array}{ccc}
c_{12}c_{13} & c_{13}s_{12} & s_{13}e^{-i\delta _{D}} \\
-c_{23}s_{12}-c_{12}s_{13}s_{23}e^{i\delta _{D}} &
c_{12}c_{23}-s_{12}s_{13}s_{23}e^{i\delta _{D}} & c_{13}s_{23} \\
s_{12}s_{23}-c_{12}c_{23}s_{13}e^{i\delta _{D}} &
-c_{12}s_{23}-c_{23}s_{12}s_{13}e^{i\delta _{D}} & c_{13}c_{23}%
\end{array}%
\right) \times U_{p},
\end{equation}%
where the Majorana phases $\theta _{\alpha,\beta }$ appear in the matrix $%
U_{p}=\mathrm{diag}(1,\,e^{i\theta _{\alpha }/2},\,e^{i\theta _{\beta }/2})$%
, and $\delta _{D}$ is the Dirac phase. The dependence on the mixing angles
is denoted by $s_{ij}\equiv \sin \theta _{ij} $, $c_{ij}\equiv \cos \theta
_{ij}$. Analysis of neutrino experimental data gives the best-fit values for
the mass-squared differences and mixing angles are $%
s_{13}^{2}=0.025_{-0.003}^{+0.003}$, $s_{23}^{2}=0.43_{-0.03}^{+0.03}$, $%
s_{12}^{2}=0.320_{-0.017}^{+0.016}$, $|\Delta
m_{13}^{2}|=2.55_{-0.09}^{+0.06}\times 10^{-3}\mathrm{eV}^{2}$, and $\Delta
m_{21}^{2}=7.62_{-0.19}^{+0.19}\times 10^{-5}\mathrm{eV}^{2}$~\cite%
{Tortola:2012te}. Matching to these values determines the regions of
parameter space with viable neutrino masses.

\section{Experimental Constraints\label{sec:constraints5}}

\begin{figure}[t]
\centering
\epsfig{file=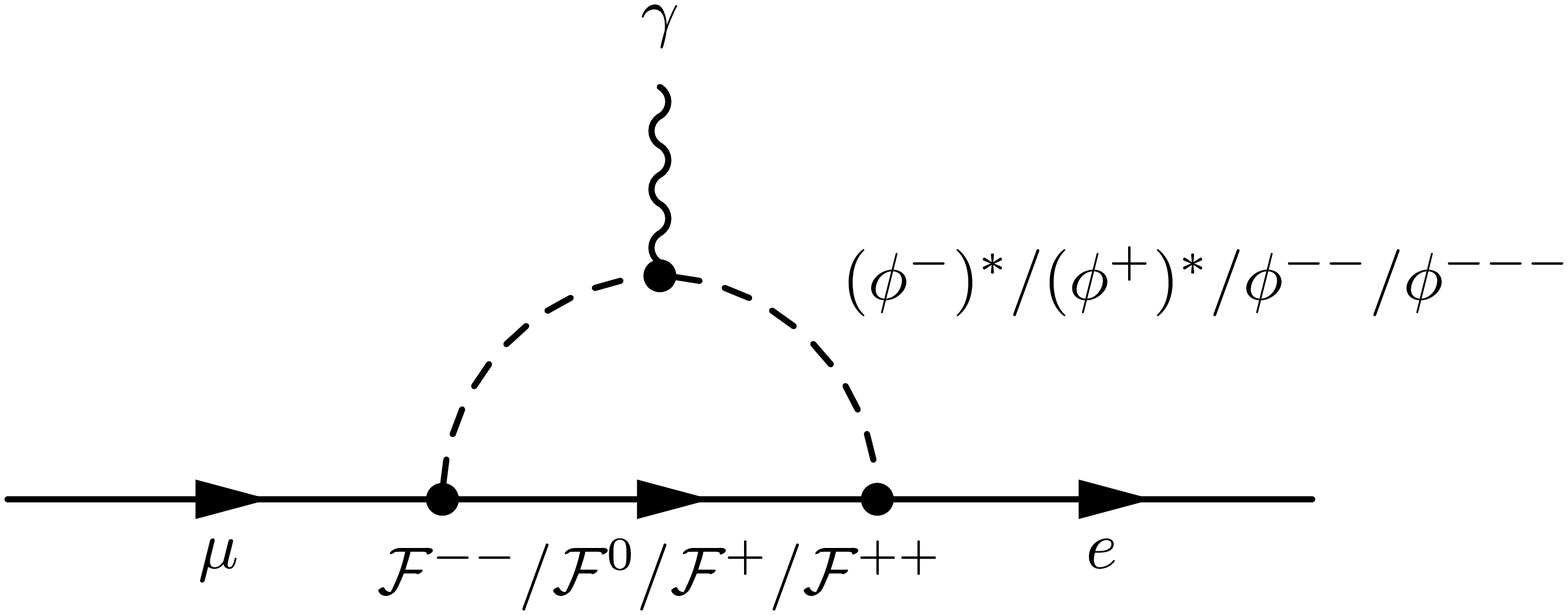,width=0.4\linewidth,clip=}~\epsfig{file=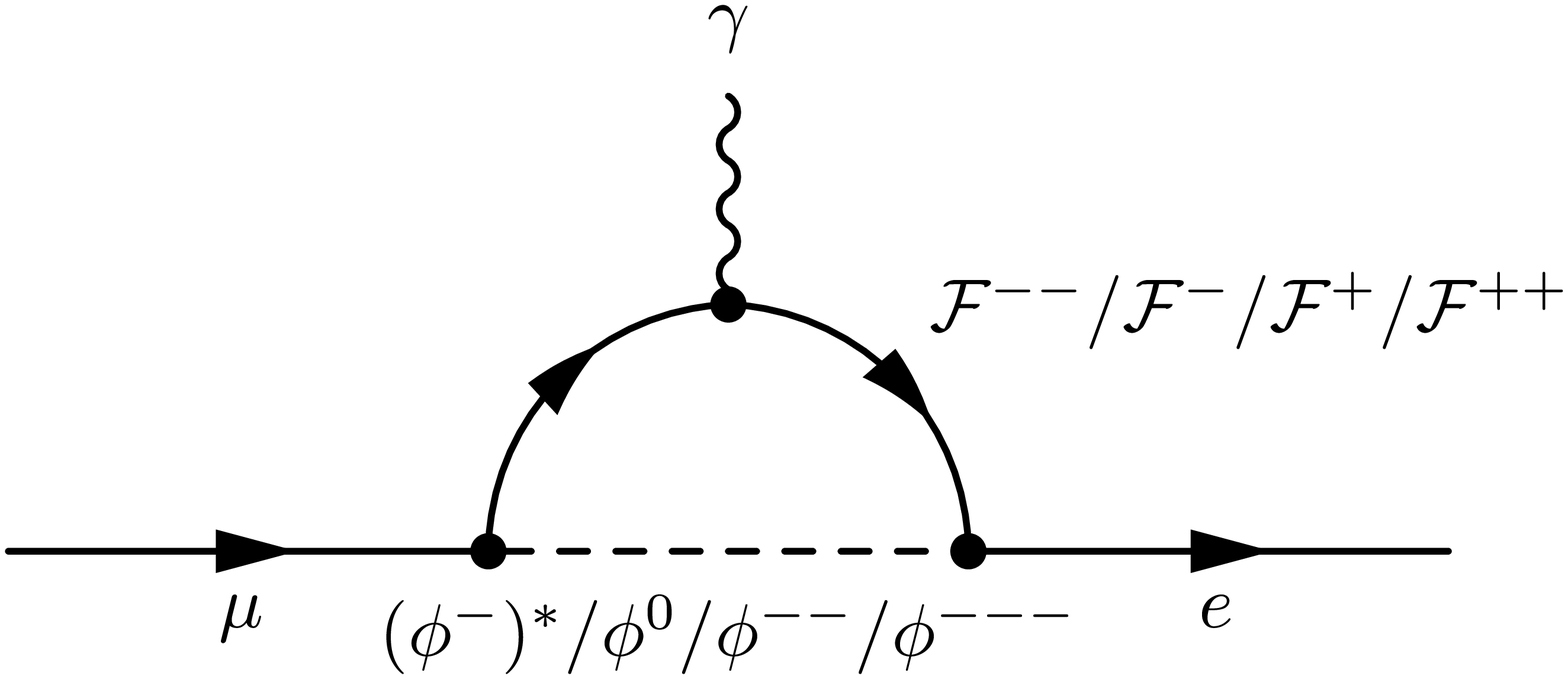,width=0.4\linewidth,clip=}
\caption{Diagrams for $\mu \rightarrow e+\gamma $ due to the
$Z_{2}$-odd fields $\mathcal{F}\sim (1,5,0)$ and $\phi \sim (1,5,2)$
. There is also a similar diagram involving the scalar $S\sim
(1,1,2)$.} \label{fig:muEgamma5}
\end{figure}

The Yukawa couplings $g_{i\alpha }$ induce flavor changing processes like $%
\mu \rightarrow e+\gamma $. At the one-loop level, there are two classes of
diagrams containing $\mathcal{F}$ and $\phi $ that one should consider, as
shown in Figure~\ref{fig:muEgamma5}. Note, however, that diagrams with the
photon attached to the internal fermion come in pairs which differ by an
overall sign. The coherent sum of the corresponding amplitudes vanishes in
the limit that the small mass-splitting are neglected.\footnote{%
Said differently, the cancelation occurs because $\sum_{\mathcal{F}}Q_{%
\mathcal{F}}=0$ for all non-trivial $SU(2)$ multiplets with vanishing
hypercharge ($Q_{\mathcal{F}}$ are the charges of the components of $%
\mathcal{F}$).} A similar cancelation occurs between the diagrams containing
singly-charged scalars in Figure~\ref{fig:muEgamma5}a. Calculating the
diagrams in Figure~\ref{fig:muEgamma5}, and adding the diagram involving the
singlet $S$, one finds that the branching fraction for $\mu \rightarrow
e+\gamma $ is given by
\begin{eqnarray}
B(\mu \rightarrow e\gamma ) &=&\frac{\Gamma (\mu \rightarrow e+\gamma )}{%
\Gamma (\mu \rightarrow e+\nu +\bar{\nu})}  \notag \\
&\simeq &\frac{\alpha \upsilon ^{4}}{384\pi }\times \left\{ \frac{|f_{\mu
\tau }f_{\tau e}^{\ast }|^{2}}{M_{S}^{4}}+\frac{900}{M_{\phi }^{4}}%
\left\vert \sum_{i}g_{ie}^{\ast }g_{i\mu }F_{2}(M_{i}^{2}/M_{\phi
}^{2})\right\vert ^{2}\right\} .  \label{eq:muEgamma_5}
\end{eqnarray}%
Here the function $F_{2}(R)=(R-1)^{-4}[1-6R+3R^{2}+2R^{3}-6R^{2}\log
R]/6$ is a standard one-loop function. A simple change of the flavor
labels in Eq.~\eqref {eq:muEgamma_5} allows one to obtain the
related expression for $B(\tau \rightarrow \mu +\gamma )$.

When the final-state electrons in Figure~\ref{fig:muEgamma5} are replaced
with muons, the diagrams contribute to the magnetic moment of the muon.
Similar arguments to those just used also apply for the calculation of the
magnetic moment; for example, the diagrams with the photon attached to the
internal-fermion line cancel in the limit the mass-splitting is neglected.
The remaining diagrams give
\begin{equation}
|\delta a_{\mu}|=\frac{m_{\mu }^{2}}{16\pi ^{2}}\left\{ \sum_{\alpha
\neq
\mu }\frac{|f_{\mu \alpha }|^{2}}{6M_{S}^{2}}+\sum_{i}\frac{5|g_{i\mu }|^{2}%
}{M_{\phi }^{2}}F_{2}(M_{i}^{2}/M_{\phi }^{2})\right\} ,
\end{equation}%
where once again a diagram involving the scalar $S$ must be included.

Null-results from searches for neutrino-less double-beta decay provide an
additional constraint of $(\mathcal{M}_{\nu })_{ee}\lesssim 0.35$~\textrm{eV}
\cite{Simkovic:2009pp} that must also be considered. We find that this
constraint is easily satisfied in the model. Next generation experiments
will improve this bound to the level of $(\mathcal{M}_{\nu })_{ee}\lesssim
0.01$~eV~\cite{Avignone:2005cs,Rodejohann:2011mu}.

\section{Dark Matter\label{sec:dark_matter5}}

\subsection{Dark Matter Longevity}

In the preceding we considered the most-general version of the model, where
all parameters allowed by the gauge symmetries are included. We showed that
the model can generate viable neutrino masses for a wide range of exotic
mass scales. In this section we turn our attention to DM, to determine
whether the model can provide a unified solution to the DM and neutrino mass
problems. The first issue to discuss is the matter of DM longevity. As noted
earlier, the model possesses a softly broken $Z_2$ symmetry, $\{\phi,%
\mathcal{F}\}\rightarrow\{-\phi,-\mathcal{F}\}$, which becomes exact in the
limit $\lambda\rightarrow0$. This suggests that the model could also give a
dark matter candidate.

There are two candidates for the DM in this model, either the scalar $\phi
^{0}$ or the lightest neutral fermion $\mathcal{F}_{1}^{0}$. However, $\phi
^{0}$ couples to the $Z$ boson and can be excluded by direct-detection
experiments, due to tree-level interactions with SM matter and an absence of
any splitting between the real and imaginary components of $\phi ^{0}$. This
leaves $\mathcal{F}_{1}$ as the sole candidate, suggesting Majorana DM and
requiring $M_{{DM}}=M_{F}<M_{\phi }$. Consider the case with $\lambda \neq 0$%
. Then, there are two types of one-loop $\mathcal{F}_{1}^{0}$ decays that
can dominate, depending on the ordering of $M_{S}$ and $M_{DM}$, namely
\begin{eqnarray}
\mathcal{F}_{1}^{0} &\longrightarrow &S+3e\quad \quad \mathrm{for}\quad
\quad M_{S}<M_{F},  \notag \\
\mathcal{F}_{1}^{0} &\longrightarrow &4e+\nu \quad \quad \mathrm{for}\quad
\quad M_{F}<M_{S}.
\end{eqnarray}%
The corresponding widths are approximately
\begin{eqnarray}
\Gamma (\mathcal{F}_{1}^{0}\rightarrow S+3e) &\sim &|\lambda |^{2}\,M_{F}\,%
\frac{|g_{1\alpha }g_{j\beta }g_{j\gamma }|^{2}}{(16\pi ^{2})^{2}}\,\Phi _{4-%
\mathrm{body}}\quad \quad \mathrm{for}\quad \quad M_{S}\ll M_{F},  \notag \\
\Gamma (\mathcal{F}_{1}^{0}\rightarrow 4e+\nu ) &\sim &|\lambda
|^{2}\,M_{F}\,\frac{|g_{1\alpha }g_{j\beta }g_{j\gamma }f_{\delta \epsilon
}|^{2}}{(16\pi ^{2})^{2}}\,\Phi _{5-\mathrm{body}}\quad \quad \mathrm{for}%
\quad \quad M_{F}<M_{S},
\end{eqnarray}%
where $\Phi _{n-\mathrm{body}}$ denotes the $n$-body phase space factor. Due
to the presence of the softly-broken accidental $Z_{2}$ symmetry, one can
always choose nonzero $\lambda \ll 1$ sufficiently small to ensure adequate
dark-matter longevity. This provides a simple way to include a DM candidate
without recourse to an additional symmetry. The limit $\lambda \rightarrow 0$
then smoothly interpolates to the $Z_{2}$-symmetric case, making $\mathcal{F}%
_{1}^{0}$ absolutely stable. Importantly, neutrino masses are not sensitive
to this limit, and viable masses can be obtained irrespective of DM
considerations.

\subsection{Relic Density}

Taking the neutral fermion $\mathcal{F}_1^0$ as the DM candidate, there are
two classes of interactions that maintain thermal contact with the SM in the
early universe. This includes processes mediated by $SU(2)_L$ gauge bosons,
which can be calculated in the $SU(2)$-symmetric limit, and others mediated
by the scalar $\phi$. One must also include coannihilation processes in the
calculation, due to the small mass-splitting between the charged and neutral
fermions.

The annihilation of DM due to $\phi $-exchange give
\begin{equation}
\sigma (2\mathcal{F}^{0}\rightarrow \ell _{\beta }^{+}\ell _{\alpha
}^{-})\times v_{r}=\frac{|g_{1\alpha }^{\ast }g_{1\beta }|^{2}}{48\pi }\frac{%
M_{F}^{2}(M_{F}^{4}+M_{\phi }^{4})}{(M_{F}^{2}+M_{\phi }^{2})^{4}}\times
v_{r}^{2}\ \equiv \ \sigma _{0,0}^{\alpha \beta }\times v_{r},
\label{eq:phi_neutral}
\end{equation}%
where $v_{r}=2v$ is the relative velocity of the dark matter in the
centre-of-mass frame. There are no $s$-wave annihilations in this expression
as the DM is a Majorana fermion and we neglected final-state lepton masses.
There are no coannihilations mediated by $\phi $, though one must include
the annihilations for singly charged fermions:
\begin{equation}
\sigma (\mathcal{F}^{-}\mathcal{F}^{+}\rightarrow \ell _{\beta }^{+}\ell
_{\alpha }^{-})\times v_{r}=\frac{|g_{1\alpha }^{\ast }g_{1\beta }|^{2}}{%
48\pi }\frac{M_{F}^{2}(M_{F}^{4}+M_{\phi }^{4})}{(M_{F}^{2}+M_{\phi
}^{2})^{4}}\times v_{r}^{2}\ \equiv \ \sigma _{-,+}^{\alpha \beta }\times
v_{r},  \label{eq:phi_charged}
\end{equation}%
and doubly-charged fermions
\begin{equation}
\sigma (\mathcal{F}^{--}\mathcal{F}^{++}\rightarrow \ell _{\beta }^{+}\ell
_{\alpha }^{-})\times v_{r}=\frac{|g_{1\alpha }^{\ast }g_{1\beta }|^{2}}{%
48\pi }\frac{M_{F}^{2}(M_{F}^{4}+M_{\phi }^{4})}{(M_{F}^{2}+M_{\phi
}^{2})^{4}}\times v_{r}^{2}\ \equiv \ \sigma _{--,++}^{\alpha \beta }\times
v_{r}.  \label{eq:phi_2charged}
\end{equation}
For annihilations and coannihilations involving $SU(2)_{L}$ gauge bosons we
can make use of known results in the literature~\cite{Cirelli:2009uv}.

In the limit where the mass-splitting between fermion components vanishes, $%
\Delta M_{\mathcal{F}}\rightarrow 0$, we add annihilation and coannihilation
channels together with the standard method~\cite{Griest:1990kh} to obtain
\begin{equation}
\sigma _{eff}(2\mathcal{F}\rightarrow SM)\times v_{r}=\frac{1}{g_{eff}^{2}}%
\left[ \sigma _{W}\times v_{r}+\sum_{\alpha ,\beta }\left\{
g_{0}^{2}\,\sigma _{0,0}^{\alpha \beta }+2g_{\pm }\,\sigma _{-,+}^{\alpha
\beta }+2g_{\pm \pm }\,\sigma _{--,++}^{\alpha \beta }\right\} \times v_{r}%
\right] ,
\end{equation}%
where the $SU(2)_{L}$ channels are denoted by
\begin{equation}
\sigma _{W}\equiv \frac{\pi \alpha _{2}^{2}}{2M_{F}^{2}v_{r}}\left\{ 2070+%
\frac{1215}{2}v_{r}^{2}\right\} ,
\end{equation}%
and $g_{eff}=g_{0}+2g_{\pm }+g_{\pm \pm }$, with $g_{0}=g_{\pm }=g_{\pm \pm
}=2$. The $\phi $-exchange cross sections are defined above.

\subsection{Direct Detection}

\begin{figure}[t]
\centering
\epsfig{file=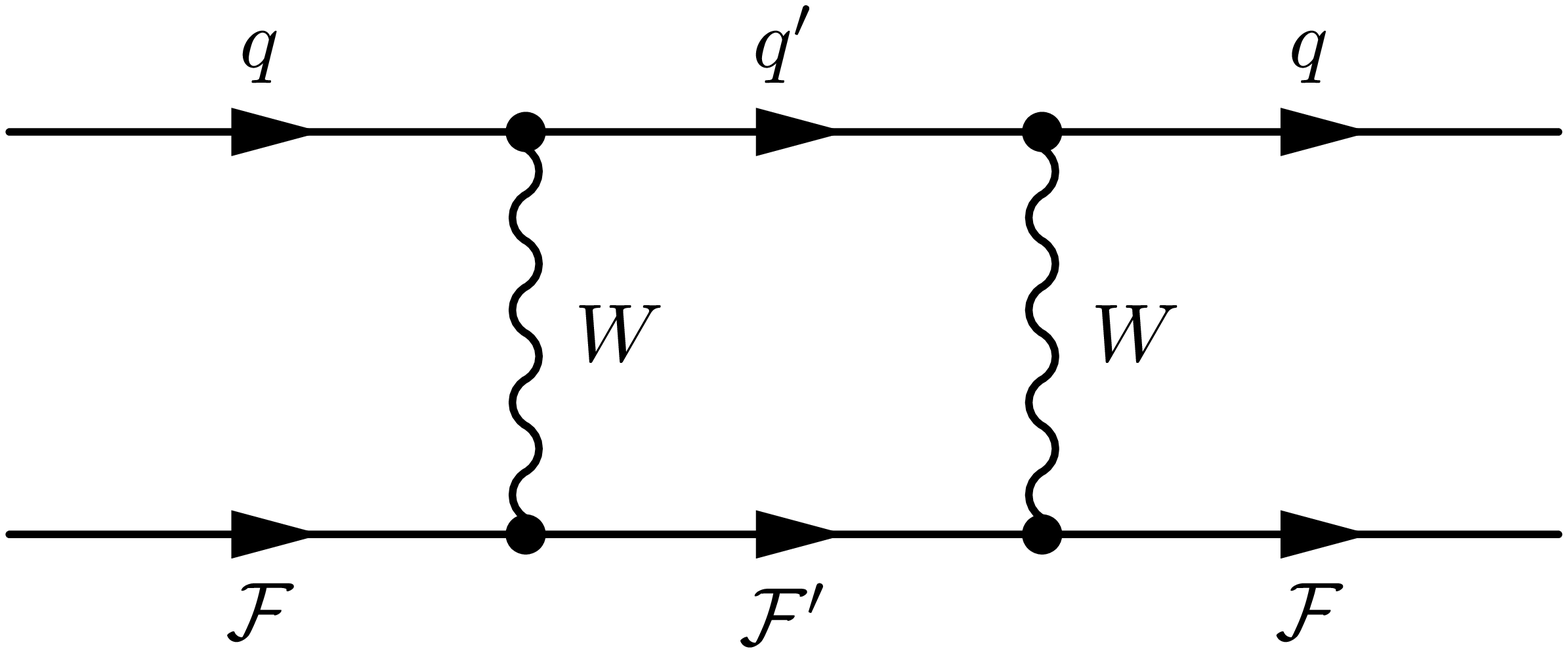,width=0.3\linewidth,clip=}~%
\epsfig{file=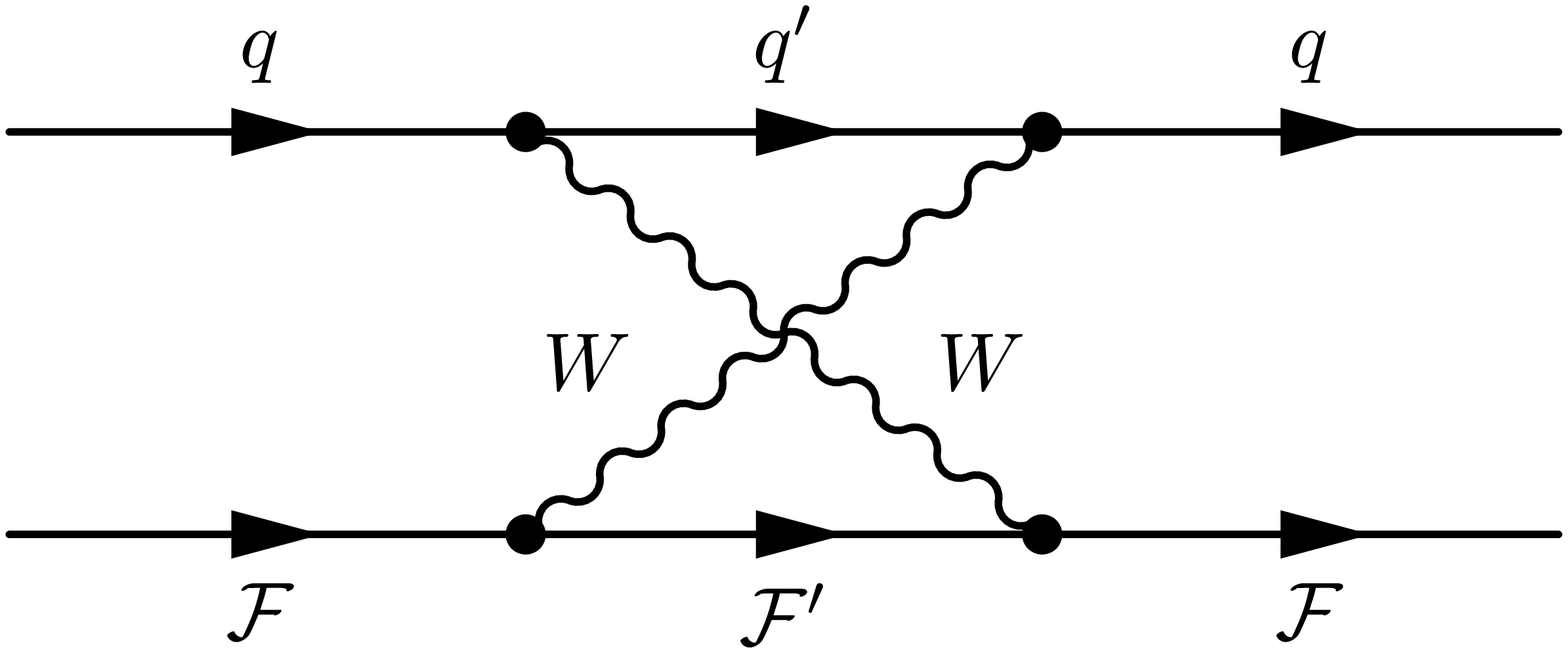,width=0.3\linewidth,clip=}~%
\epsfig{file=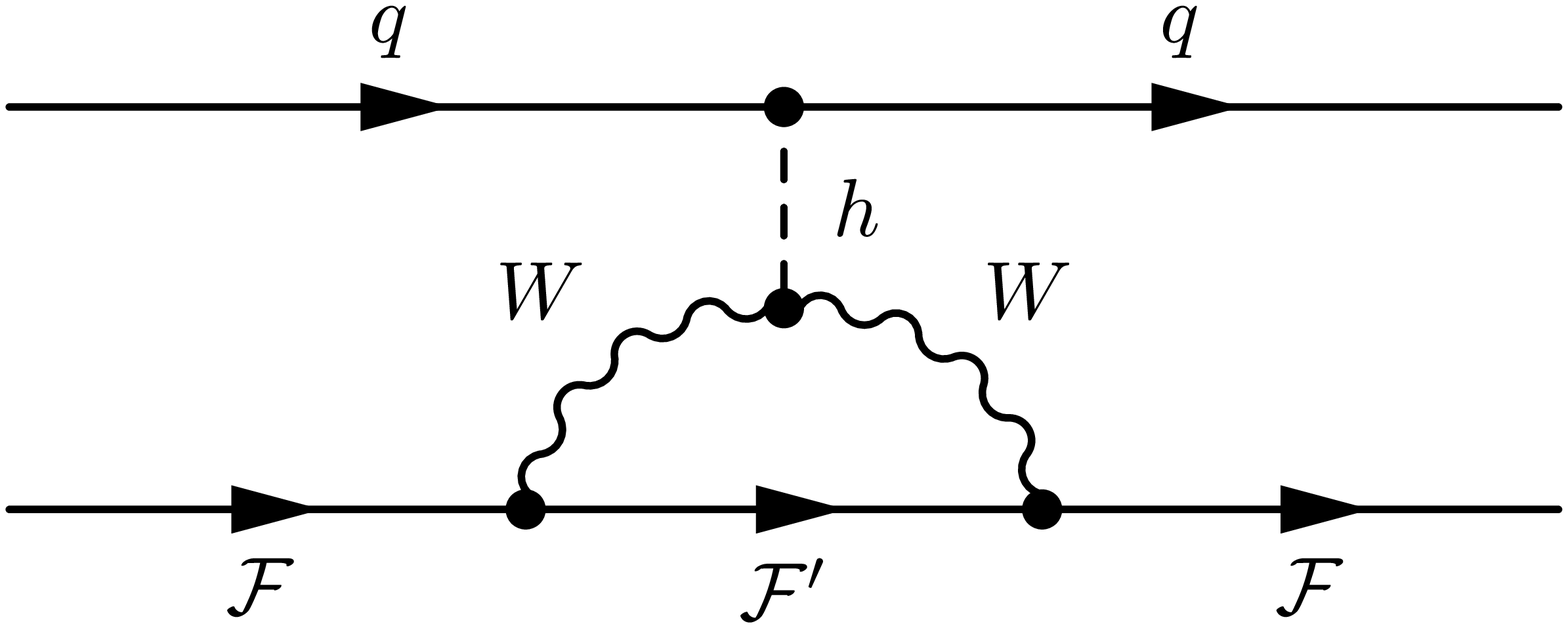,width=0.3\linewidth,clip=}
\caption{Feynman diagrams for direct-detection experiments.}
\label{fig:DD_figure_5plet}
\end{figure}

The DM candidate does not couple to quarks at tree-level due to its
vanishing hypercharge and, being a Majorana fermion, there are no radiative
magnetic-dipole interactions with SM gauge bosons. Exchange of $W$ bosons
generates the three one-loop diagrams in Figure~\ref{fig:DD_figure_5plet},
which are relevant for direct-detection experiments. The scattering contains
both spin-dependent and spin-independent contributions. However, the former
are suppressed by the DM mass, expected to be $M_{F}\sim 10$~TeV in our
case, giving highly-suppressed spin-dependent scattering cross sections.

The dominant interaction is therefore spin-independent scattering, with a
cross section determined by SM interactions:
\begin{equation}
\sigma _{\mathrm{SI}}(\mathcal{F}^{0}N\rightarrow \mathcal{F}^{0}N)\simeq
\frac{9\pi \alpha _{2}^{4}M_{A}^{4}f^{2}}{M_{W}^{2}}\left[ \frac{1}{M_{W}^{2}%
}+\frac{1}{M_{h}^{2}}\right] ^{2}.
\end{equation}%
Here the DM scatters off a target nucleus $A$ with mass $M_{A}$ and we use a
standard parameterizations for the nucleon,
\begin{equation}
\langle N|\sum_{q}m_{q}\,\bar{q}q\,|N\rangle \ =\ f\,m_{N},
\end{equation}
with $m_{N}$ being the nucleon mass. We use $f\approx 1/3$, though this is
subject to the standard QCD uncertainties. The resulting cross section per
nucleon is of order $\sigma _{\mathrm{SI}}\simeq 10^{-46}\mathrm{cm}^{2}$,
which is beyond the current sensitivity of LUX \cite{Akerib:2013tjd}, but
within reach of forthcoming experiments like SuperCDMS \cite{superCDMS}.
Discovery prospects are therefore promising.

\section{Results and Discussion\label{sec:results}}

\begin{figure}[t]
\centering \epsfig{file=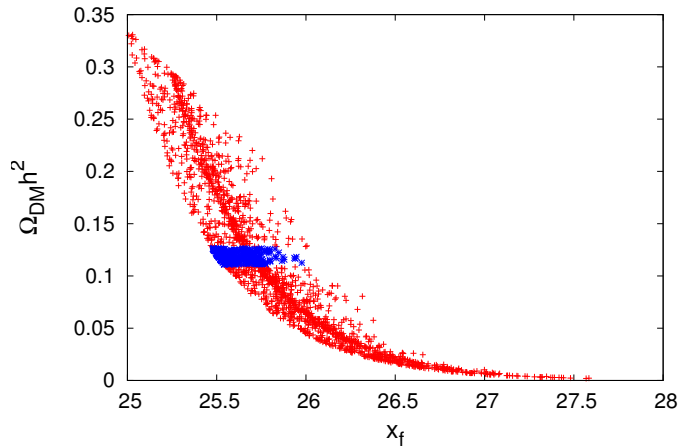,width=0.6\linewidth,clip=}
\caption{The DM relic abundance versus the (scaled) DM mass, $%
x_{f}=M_{F}/T_{f}$, where the blue benchmarks correspond to the physical
observed value.}
\label{Omega}
\end{figure}

\begin{figure}[h]
\centering
\epsfig{file=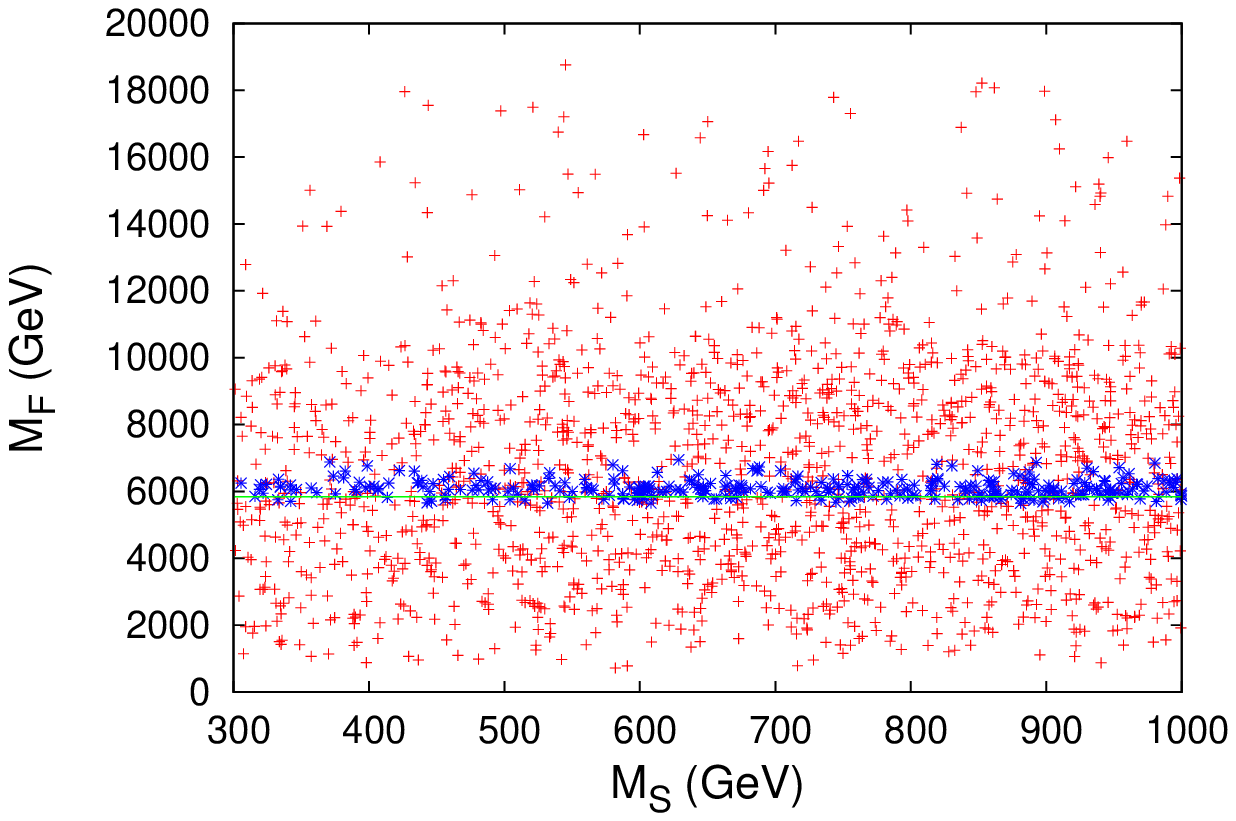,width=0.5\linewidth,clip=}~%
\epsfig{file=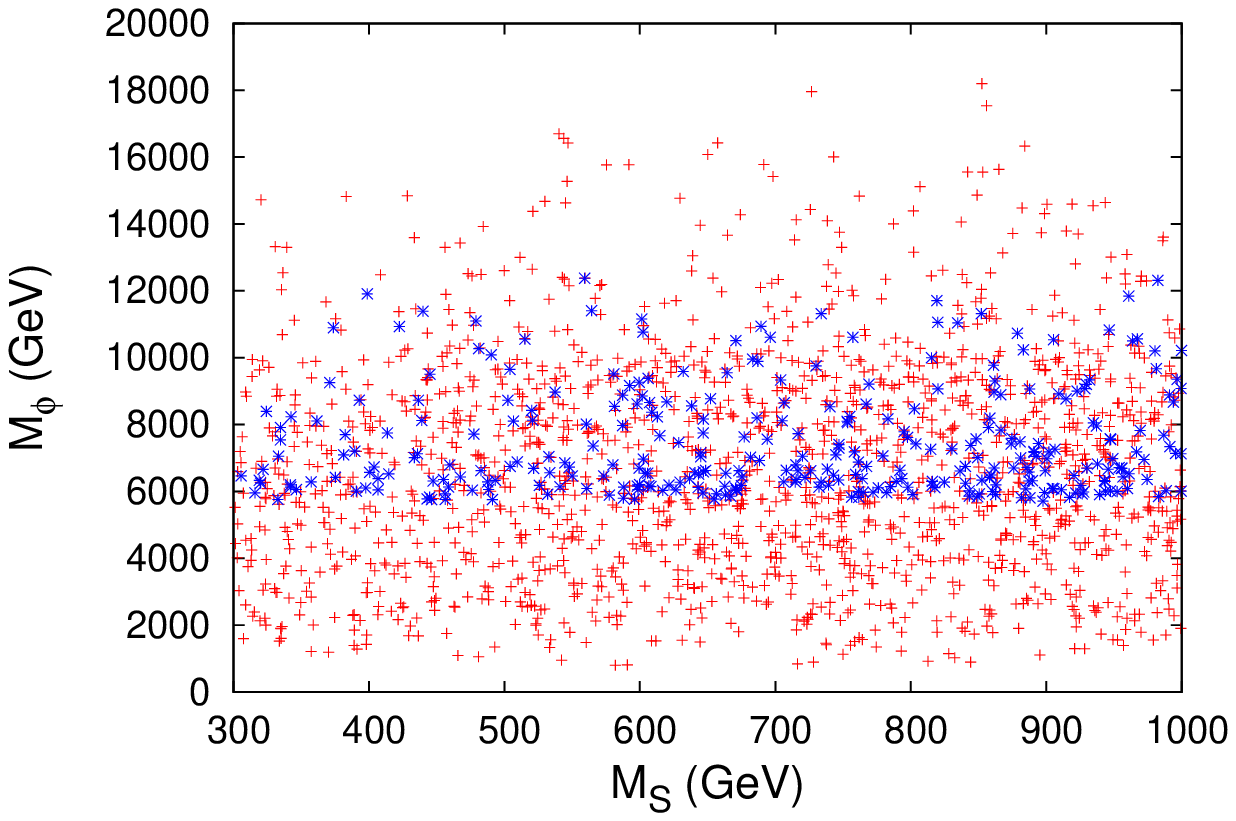,width=0.5\linewidth,clip=} \caption{Allowed
mass values. These achieve viable neutrino mass/mixings while
satisfying the constraints. The blue points give the DM relic
abundance in accordance with Figure~\ref{Omega}. Left: The lightest
neutral-fermion mass versus the singlet scalar mass. When the
correct relic abundance is achieved, $\mathcal{F}_{1}$ is the DM,
with $M_{F}$. The green
line gives the best-fit value for $\Omega_{DM}h^{2}$ when $g_{i\alpha%
}\rightarrow 0$. Right: The corresponding scalar masses, with $M_{F}<M_{%
\phi}$ assumed.} \label{fig:masses}
\end{figure}

\begin{figure}[h]
\centering
\epsfig{file=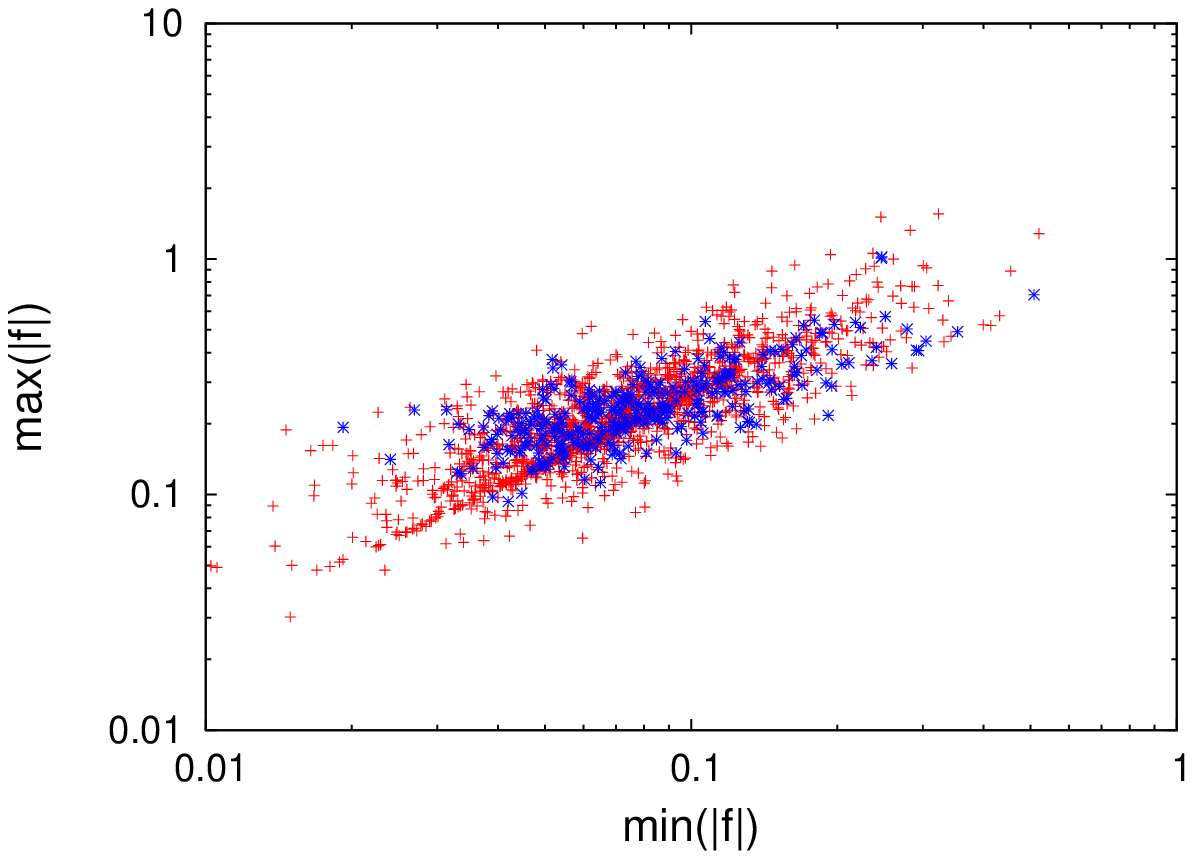,width=0.5\linewidth,clip=}~\epsfig{file=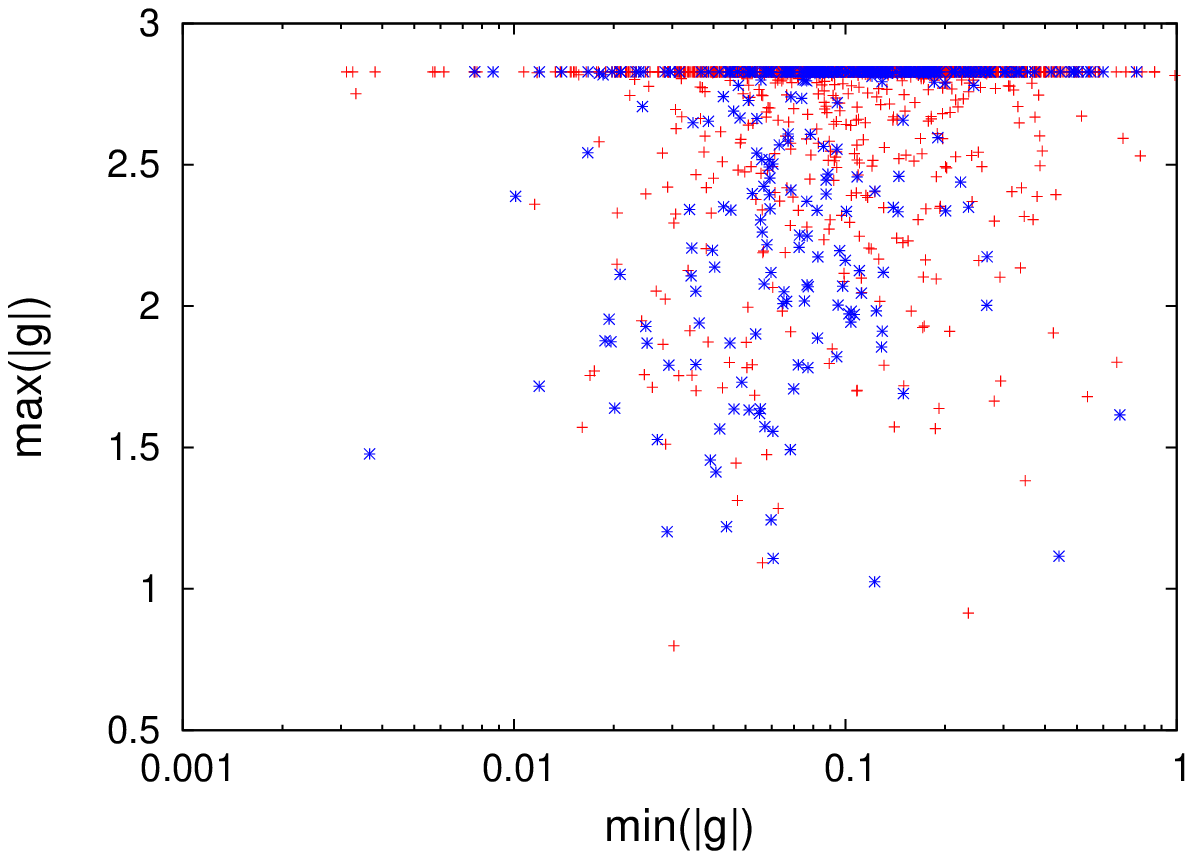,width=0.5%
\linewidth,clip=}
\caption{Viable regions of parameter space for the Yukawa couplings $f_{%
\alpha\beta}$ and $g_{i\alpha}$. Correct neutrino mass and mixing is
obtained and flavor-changing constraints are satisfied.
The blue benchmarks give the DM relic abundance in accordance with Figure~%
\ref{Omega}.} \label{fig:couplings}
\end{figure}

\begin{figure}[h]
\centering \epsfig{file=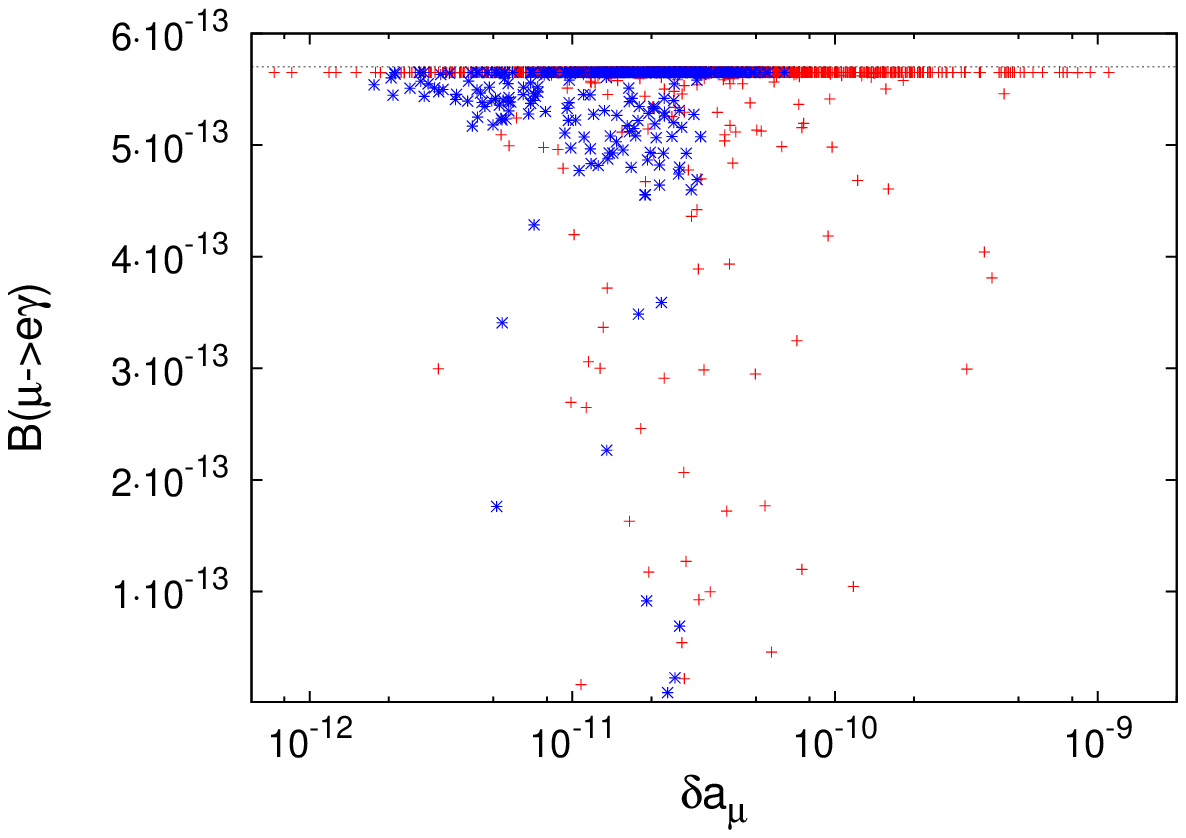,width=0.5\linewidth,clip=}~ %
\epsfig{file=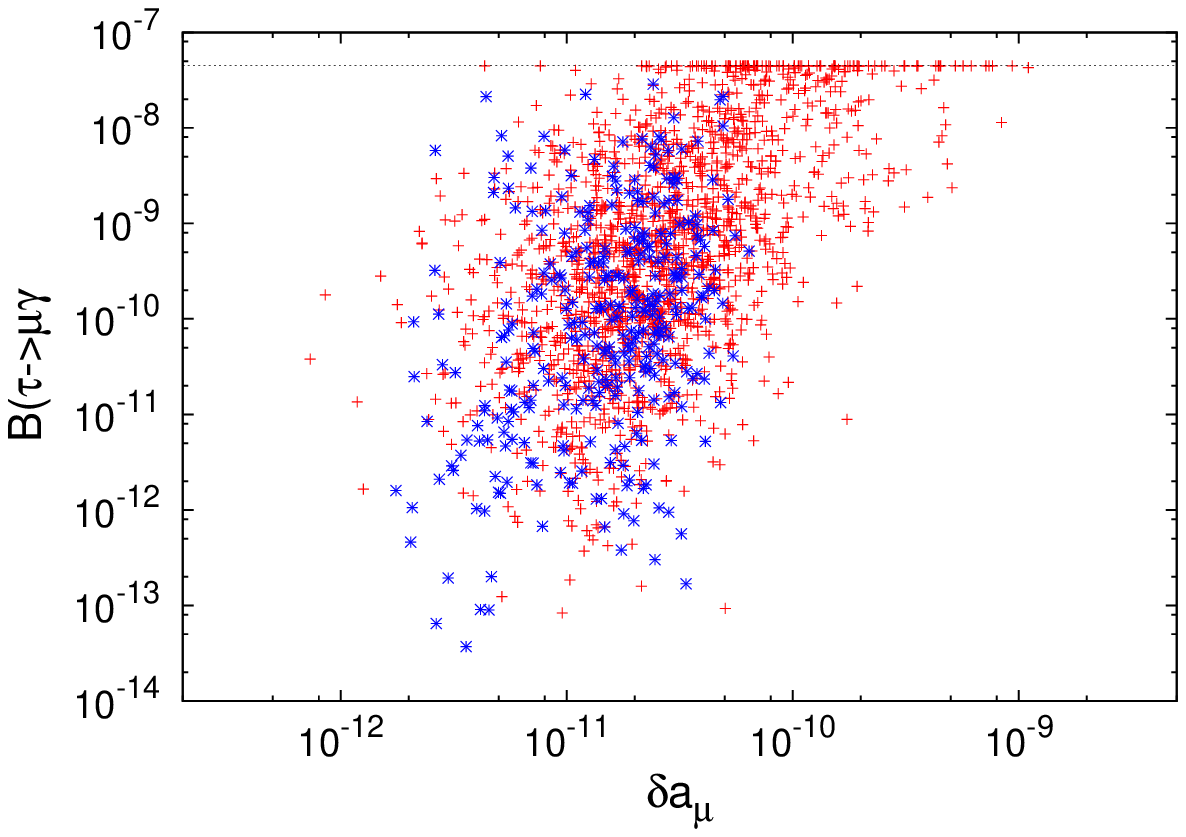,width=0.5\linewidth,clip=}
\caption{Branching fractions for lepton flavor violating decays versus the
anomalous magnetic moment of the muon. The dashed lines represent the
experimental upper bounds on the branching ratios.}
\label{fig:muEgamma}
\end{figure}

Using the results from the preceding sections, we can determine the
parameter space where viable neutrino masses are obtained and the correct DM
relic-density is realized. Here, we present the results from our numerical
scans of the parameter space. We find that neutrino masses can be obtained
for a range of parameter space, including the fermion and scalar masses. It
also appears that the observed DM relic abundance can be generated. Whenever
we consider $\mathcal{F}_{1}$ as DM, we assume $\lambda $ is sufficiently
small to ensure DM longevity. In our numerical scan, we consider the
following range for the model parameters
\begin{eqnarray*}
~\left\vert f_{\alpha \beta }\right\vert ^{2},\,\left\vert g_{i\alpha
}\right\vert ^{2}\lesssim 9,\quad~500~\mathrm{GeV}\leq M_{F}\leq 10~\mathrm{%
TeV}, \\
300~\mathrm{GeV}\leq M_{S}\leq 1~\mathrm{TeV},\quad~M_{2,3},M_{\phi }\gtrsim
M_{F},
\end{eqnarray*}%
and we impose the constraints from neutrino mass and mixing, LFV processes
and muon anomalous magnetic moment, with and without the DM relic abundance
constraints. In Figures \ref{Omega}, \ref{fig:masses}, \ref{fig:couplings}
and \ref{fig:muEgamma}, the red (bule) benchmarks represent the sets of
model parameters that satisfy the constraints without (with) the DM relic
density abundance.

In Figure~\ref{Omega} we plot the relic density, $\Omega _{DM}h^{2}$, versus
the (scaled) DM mass, $x_{f}=M_{F}/T_{f}$, where $T_{f}$ is the freeze-out
temperature. The blue points correctly reproduce the observed relic-density~%
\cite{Ade:2013ktc}. Viable neutrino masses are obtained for all points shown
and the various constraints are satisfied; regions of parameter space that
do not give the observed relic-abundance still allow a viable model of
neutrino mass. For parameter space where the DM abundance is too large one
must take $\lambda $ adequately large to allow $\mathcal{F}_{1}$ to decay to
the SM. In other cases one can consider small finite values of $\lambda $ or
simply take $\lambda \rightarrow 0$ to achieve the $Z_{2}$-symmetric limit.

The corresponding masses for the exotic fields are shown in Figure~\ref%
{fig:masses}. In the limit where the annihilations involving $\phi $ are
switched off, $g_{i\alpha }\rightarrow 0$, the green line in Figure ($%
M_{F}=5.844$ $\mathrm{TeV}$)~\ref{fig:masses}-left corresponds to the
current best-fit value for the DM relic density, $\Omega _{DM}h^{2}=0.1187$.
We observe that $\phi $-exchange slightly modifies the value of $M_{F}$ by a
ratio between [-3.3\%,19\%] with the mass $M_{F}\sim 6$ $TeV$ generically
expected. When DM is incorporated, the fermions $\mathcal{F}$ and the scalar
$\phi $ are both well-beyond the reach of collider experiments. On the other
hand, taken purely as a model of neutrino mass, these exotics can have $%
\mathcal{O}(\mathrm{TeV})$ masses as seen in Figure~\ref{fig:masses}. In
either case, the singlet scalar $S$ can remain relatively light with mass $%
M_{S}=\mathcal{O}(100)$~GeV, so collider experiments should provide
additional tests on the model. In our analysis we restricted our scans to
parameter space with $M_{F}<M_{\phi }$, as required when $\mathcal{F}_{1}$
is the DM. When only neutrino masses are considered one could consider
alternative mass orderings for the exotics, with $M_{\phi }<M_{\mathcal{F}}$
also possible.

The viable parameter space for the Yukawa couplings $f_{\alpha \beta }$ and $%
g_{i\alpha }$ is shown in Figure~\ref{fig:couplings}. In our numerical scans
we restricted these couplings to the perturbative range, $\left\vert
f_{\alpha \beta }\right\vert ^{2},\,\left\vert g_{i\alpha }\right\vert
^{2}\lesssim 4\pi $. A reasonable spread of values are possible for $%
f_{\alpha \beta }$, though the scans generically require $g_{i\alpha }=%
\mathcal{O}(1)$. The corresponding branching fractions for the
flavor-changing decays appear in Figure~\ref{fig:muEgamma}. The bound on $%
\tau \rightarrow \mu +\gamma $ is easily satisfied, though the constraint of
$B(\mu \rightarrow e+\gamma )<5.7\times 10^{-13}$~\cite{Adam:2013mnn} makes
the parameter space very constrained. An order of magnitude improvement in
the bound on $B(\mu \rightarrow e+\gamma )$ would exclude the vast majority
of the viable parameter space found in the scans. It is worth noting that
with only two generations of fermions $\mathcal{F}_{i}$ ($g_{3\alpha}=0$,
the bound on $B(\mu \rightarrow e+\gamma )$ is violated. Three generations
of $\mathcal{F}_{i}$ are therefore required to obtain agreement with
constraints from lepton flavor violating processes.

In the above we employed the DM annihilation cross sections from Section~\ref%
{sec:dark_matter5}, finding $M_{F}\sim 6$~TeV. With this value of $M_{F}$,
low-energy constraints are readily satisfied and viable neutrino masses are
obtained. However, the (co-)annihilation cross sections are subject to a
Sommerfeld enhancement due to $SU(2)_{L}$ gauge-boson exchange. This
modifies the (co-)annihilation cross sections and increases $M_{F}$. When
the enhancement is applied to $s$-wave (co-)annihilations via $SU(2)_{L}$
interactions, the requisite DM mass increases to $M_{F}\sim 10$~TeV~\cite%
{Cirelli:2005uq}. In our case one should also include the enhancement for
the $p$-wave annihilations~\cite{Cassel:2009wt}, which is beyond the scope
of this work. However, we anticipate similar results for our model and
expect an $\mathcal{O}(1)$ correction to $M_{F}$ due the enhancement. To
determine if the model is likely to remain viable once the Sommerfeld effect
is included, we studied the parameter space with heavier $M_{F}\lesssim 20$%
~TeV. We found that viable neutrino mass/mixings could be obtained while
satisfying the various constraints. These results are already incorporated
in the figures, as seen in Figure~\ref{fig:masses}, where $M_{F}\lesssim 20$%
~TeV is considered. These results indicate that the model should remain
viable with $M_{F}\sim 10$~TeV.

\section{Collider Phenomenology\label{sec:collider}}

Although a detailed study of the collider phenomenology of our model
is beyond the scope of the present work, we briefly discuss some
important signatures at both the LHC and future $e^+e^-$ colliders.
If $\mathcal{F}_1$ provides the DM relic abundance one requires
$M_f\sim10$~TeV with $M_\phi>M_F $, placing both $\mathcal{F}$ and
$\phi$ well beyond the reach of foreseeable collider experiments.
However, the singlet charged scalar $S^\pm$ can remain within reach
of TeV scale colliders. At the International Linear collider (ILC)
\cite{ILC}, the charged scalars $S^\pm$ can be directly produced
through the t-channel process $e^+e^-\rightarrow S^+ S^- \rightarrow
l_{\alpha}^+ l_{\beta}^- + E_{miss}$, which includes lepton flavor
violating final-states that can be observed as a pair of charged
leptons with missing energy (similar to the KNT model
\cite{Ahriche:2014xra}). However, due to different constraints in
the KNT model, the corresponding charged scalar is not allowed to be
as light as 300 GeV, as is the case here. Therefore it should be
easier to test our model through this channel at the ILC for
energies $\sqrt{s}=500$~GeV and 1 TeV. At the LHC, this model can
similarly be probed via the process $pp\rightarrow S^+ S^-
\rightarrow l_{\alpha}^+ l_{\beta}^- + E_{miss}$ with the charged
scalars produced through Drell-Yan.

The region of parameter space with lighter values of $M_F\sim$~TeV is also
of interest as it allows $\mathcal{F}$ (and possibly $\phi$) to be within
reach of collider experiments like the LHC. In this parameter space $%
\mathcal{F}_1$ cannot provide the full  DM relic abundance though it can
provide a sub-leading contribution. The exotic fermions would be pair
produced via weak interactions at the LHC as $pp\rightarrow W/Z\rightarrow
\mathcal{F}\mathcal{F}$, with typical weak-scale cross sections (e.g.~for $%
M_F\approx 300$~GeV, one expects a production cross section of $\mathcal{O}%
(10^2)$~fb at the $7$~TeV LHC and $\mathcal{O}(10^3)$~fb at a 14~TeV LHC).
Due to the exact (or approximate) $Z_2$ symmetry, the heavier fermions must
decay weakly to lighter exotic fermions rather than directly to SM
particles. For example, one could have the production process $pp\rightarrow
W^+\rightarrow \mathcal{F}^{++}\mathcal{F}^-$, with the charged fermions
decaying via off-shell $W$ bosons to leptonic final states as $\mathcal{F}%
^{++}\rightarrow \mathcal{F}^+ \ell^+ \nu_\ell$, and $\mathcal{F}%
^\pm\rightarrow \mathcal{F}^0 \ell^\pm \nu_\ell$, where $\ell=e,\mu,\tau$
denotes the SM lepton flavor. A typical final state would contain three
charged leptons and missing energy, due to the DM and the neutrinos. Related
final states with four charged leptons are also possible via $\mathcal{F}%
^{++}\mathcal{F}^{--}$ pair production.

\section{Generalized KNT Models\label{sec:generalize_KNT}}

The model presented here is related to the proposal of KNT~\cite%
{Krauss:2002px} and a recently discovered three-loop model with triplet
fermions~\cite{Ahriche:2014cda}. In this section we identify this
relationship and show that the models form a larger set of generalized KNT
models. Consider the loop diagram in Figure~\ref{fig:3loop_nuDM5}. Adding $S$
to the SM to allow the outer vertices, the choice for $\mathcal{F}$ and $%
\phi $ is not unique. One can determine the basic conditions for a general
fermion $\mathcal{F}\sim(1,R_\mathcal{F},Y_\mathcal{F})$ and scalar $%
\phi\sim(1,R_\phi,Y_\phi)$ that allow Figure~\ref{fig:3loop_nuDM5} to
appear. The top vertex in Figure~\ref{fig:3loop_nuDM5} requires a term $%
\lambda_s(S^-)^2\phi^2\subset V(H,\,\phi,\,S)$ in the potential. This fixes $%
Y_\phi=Y_S=2$, which in turn fixes $Y_\mathcal{F}=-(Y_\phi+Y_{e_R})=0$. For
even-valued $R_\mathcal{F}$, the model contains fractionally charged
particles, the lightest of which is automatically stable and therefore
excluded by cosmological constraints. Consequently only odd-valued $R_%
\mathcal{F}$ is viable, giving $R_F=(2n+1)$ for $n=0,1,\ldots$ The $%
\overline{\mathcal{F}}\phi e_R$ vertex then fixes $R_\phi=R_\mathcal{F}%
=(2n+1)$.

For $n=0$ one has the KNT model, with $\mathcal{F}\sim(1,1,0)$ and $%
\phi\sim(1,1,2)$~\cite{Krauss:2002px}, while $n=1$ gives the
recently-proposed triplet model with $\mathcal{F}\sim(1,3,0)$ and $%
\phi\sim(1,3,2)$~\cite{Ahriche:2014cda}. In both of these models one
requires a new symmetry to remove the tree-level seesaw contributions. For $%
n=2$ one obtains the present model, which gives $\mathcal{F}\sim(1,5,0)$ and
$\phi\sim(1,5,2)$. Thus, $n=2$ is the smallest value for which no symmetry
is required to remove a tree-level seesaw mass --- neutrino mass
automatically appears at the three-loop level for $n\ge2$, irrespective of
whether a $Z_2$ symmetry is imposed.

We saw that DM longevity did not require a $Z_{2}$ symmetry in the $n=2$
model due to the softly-broken accidental $Z_{2}$ symmetry (which becomes
exact for $\lambda \rightarrow 0$). This feature is common for all
even-valued $n$ with $n\geq 2$, which is seen as follows. For all $n\geq 0$,
the most-general Lagrangian seemingly contains the term $\lambda (S^{-})\phi
^{\ast }\times (\phi \times \phi )_{R_{\mathcal{F}}}\subset V(H,\,S,\,\phi )$%
, which breaks the $Z_{2}$ symmetry. Here $(\phi \times \phi )_{R_{\mathcal{F%
}}}$ denotes the $SU(2)$-contraction of $\phi \times \phi $ in the $R_{%
\mathcal{F}}$ representation; for odd-valued $R_{\mathcal{F}}$ this is
always contained in the $SU(2)$-product, $R_{\mathcal{F}}\subset R_{\mathcal{%
F}}\times R_{\mathcal{F}}$. For $n<2$, however, the models contain
additional $Z_{2}$ symmetry breaking terms, including some that generate
tree-level neutrino masses. On the other hand, for $n\geq 2$ the $\lambda $%
-term is the sole $Z_{2}$ symmetry breaking term. Thus, $n=2$ marks the
transition where the $\lambda $-term softly breaks the $Z_{2}$ symmetry, and
all models with $n\geq 2$ seemingly possess a softly-broken accidental
symmetry that becomes exact in the limit $\lambda \rightarrow 0$. However,
although group theory gives $R_{\mathcal{F}}\times R_{\mathcal{F}}\supset R_{%
\mathcal{F}}$, the product $(\phi \times \phi )_{R_{\mathcal{F}}}$ in fact
vanishes when the scalar is in the $R_{\mathcal{F}}=2n+1$ representation for
odd-valued $n$.\footnote{%
For distinct  scalars $\phi$ and $\phi^{\prime }$, both in the $R_{\mathcal{F%
}}=2n+1$ representation, the $SU(2)$ product $(\phi \times \phi )_{R_{%
\mathcal{F}}}$ is nonzero for all $n$. However, for identical scalars $%
\phi=\phi^{\prime }$, one finds $(\phi \times \phi )_{R_{\mathcal{F}}}=0$
for odd-valued $n$.} Thus, for all even-valued $n\ge2$, the models contain
an accidental $Z_2$ symmetry that is softly broken by the term $\lambda
(S^{-})\phi ^{\ast }\times (\phi \times \phi )_{R_{\mathcal{F}}}\subset
V(H,\,S,\,\phi )$.

There is a very interesting by-product of these observations. If $\phi $ is
in the $R_{\mathcal{F}}=2n+1$ representation with $n\ge2$, the $\lambda$%
-term is the sole $Z_2$ symmetry breaking term in the model. However, for
odd-valued $n$, the $\lambda$-term vanishes identically, and the accidental $%
Z_2$ symmetry becomes an exact symmetry of the full Lagrangian. Thus, for $%
R_{\mathcal{F}}=7$, corresponding to $\mathcal{F}\sim(1,7,0)$ and $%
\phi\sim(1,7,2)$, one automatically obtains a model of radiative neutrino
mass with a stable DM candidate due to an exact accidental $Z_2$ symmetry
--- no additional symmetry need be imposed.\footnote{%
We thank T.~Toma for communications on this point.} More generally, models
with odd-valued $n>2$ will generate neutrino mass and give stable DM
candidates without invoking new symmetries.

\section{Conclusion\label{sec:conc5}}

We presented a three-loop model of neutrino mass whose most-general
Lagrangian contains a softly-broken accidental $Z_{2}$ symmetry. In the
limit that a single parameter vanishes, $\lambda \rightarrow 0$, the $Z_{2}$
symmetry becomes exact and the model contains a stable DM candidate. Even
for nonzero $\lambda \ll 1$, however, the model can give a long-lived DM
candidate. The model is related to the KNT model and its triplet variant,
with the $Z_{2}$ symmetry being equivalent to the symmetry imposed in those
models. In the present case, though, the symmetry is not needed to preclude
tree-level neutrino mass, giving a viable model of neutrino mass
irrespective of DM considerations. For sufficiently small $\lambda $, the
model gives a unified solution to the DM and neutrino mass problems, with
the novel feature of not requiring that a symmetry be imposed. We showed
that neutrino mass can be generated and that important flavor-changing
constraints can be satisfied. Taken purely as a neutrino mass model, the new
physics can be $\mathcal{O}(\mathrm{TeV})$, allowing the model to be
explored at colliders. However, when DM is included the quintuplet fields
must be heavy, with $M_{F}\sim 10$~TeV, so that only the singlet scalar $S$
can be within reach of colliders. None the less, the DM can be tested in
future direct-detection experiments. We also noted interesting
generalizations of this model in which DM stability results from an exact
accidental symmetry, the simplest of which uses septuplet $SU(2)$ fields
instead of quintuplets.

\section*{Acknowledgments\label{sec:ackn}}

The authors thank T.~Toma. AA is supported by the Algerian Ministry of
Higher Education and Scientific Research under the CNEPRU Project No
D01720130042. KM is supported by the Australian Research Council.

\appendix

\section{Radiative Neutrino Mass\label{app:loop_integral}}

The Majorana neutrino masses are calculated to be
\begin{equation}
(\mathcal{M}_{\nu })_{\alpha \beta }=\frac{5\lambda _{S}}{(4\pi ^{2})^{3}}%
\frac{m_{\gamma }m_{\delta }}{M_{\phi }}\,f_{\alpha \gamma
}\,f_{\beta \delta }\,g_{\gamma i}^{\ast }\,g_{\delta i}^{\ast
}\times F\left( \frac{ M_{i}^{2}}{M_{\phi
}^{2}},\frac{M_{S}^{2}}{M_{\phi }^{2}}\right) ,
\end{equation}%
where
\begin{equation}
F(\alpha ,\beta )=\frac{\sqrt{\alpha }}{8\beta ^{2}}\int_{0}^{\infty }dr%
\frac{r}{r+\alpha }\left( \int_{0}^{1}dx\ln \frac{x(1-x)r+(1-x)\beta +x}{%
x(1-x)r+x}\right) ^{2}.
\end{equation}%
In obtaining this form of $F$ we have neglected the lepton masses.

\end{document}